\documentclass[prl,floatfix,twocolumn,superscriptaddress,showpacs]{revtex4}

\usepackage{textcomp}
\usepackage{setspace}
\usepackage{tipa}
\usepackage{verbatim} 
\usepackage{amsmath}
\usepackage{bbold}
\usepackage{subfigure}
\usepackage{amssymb}
\usepackage{bbm}
\usepackage{accents}
\usepackage{amscd}
\usepackage{graphicx}
\usepackage{epstopdf}
\usepackage{caption}
\usepackage{subfig}
\usepackage{pdfpages}
\usepackage{ dsfont }
\usepackage{multirow}

\def\aa{\hat{a}}
\def\ad{\hat{a}^{\dag}}

\def\be{\begin{equation}}
\def\ee{\end{equation}}
\def\bea{\begin{eqnarray}}
\def\eea{\end{eqnarray}}

\def\eta{{\it et. al.}}

\def\haf{\mbox{Haf}}
\def\lhaf{\mbox{Lhaf}}

\def\one{\bar{\mathbbm{1}}}

\newcommand{\etal}{\textit{et al.}}

\definecolor{darkgreen}{rgb}{0,0.4,0}

\newcommand{\prague}{FNSPE, Czech Technical University in Prague, Br\^ehov\'{a} 7, 119 15, Praha 1, Czech Republic}

\date{\today}

\begin{document}

\title{Boson Sampling from Non-Gaussian States}
\author{Craig S. Hamilton}
\email{hamilcra@fjfi.cvut.cz}
\affiliation{\prague}
\author{Igor Jex}
\affiliation{\prague}

\begin{abstract}

Boson sampling has emerged as an important tool to demonstrate the difference between quantum and classical computers and has attracted the interest of experimentalists and theoreticians. 
In this work we study Boson sampling from general, single-mode states using a scheme that can generate any such state by combining Gaussian states and photon number measurements. We derive a formula that can be used to calculate the output photon number probabilities of these states after they travel through a linear interferometer. 
This extends the Boson sampling protocol to the widest array of possible single-mode states and from this we  show that the complexity scaling of all such states is similar and hence there is no complexity advantage of using complex input states over simpler ones.

\end{abstract}


\maketitle

\section{Introduction}

Demonstrations of quantum advantage are important tests of noisy intermediate-scale quantum computing (NIQC) \cite{Bharti:2022p23008}, with quantum sampling problems emerging as promising milestones in this pursuit \cite{Hangleiter:2022p23584}. 
The aim of these sampling protocols is to use quantum mechanics to create a large quantum state, whose measurement outcomes are computationally difficult for a classical computer to recreate. 
Within sampling problems, Boson sampling has established itself as an important example in this class \cite{Lund:2017p14714, Brod:2019p23599}.

In the original Boson sampling protocol, devised by Aaronson and Arkhipov \cite{Aaronson:2013p7598}, single photons are sent into the modes of a linear interferometer and the output state is then measured in the photon number basis.  
The probability of  various output states can be written in terms of the permanent of a matrix, which itself is derived from the unitary matrix that describes the interferometer. 
The permanent is in the \#P-hard complexity class \cite{arora2009computational}  and is thus classically intractable to calculate for large matrices. Thus the probability distribution can be sampled more efficiently from this quantum device than a classical one. 
The authors show that if there was a classical algorithm that could efficiently approximate the permanent to reasonable error margins, then this would have such significant implications for complexity theory it is not believed to be true. 
The original version of Boson sampling has been experimentally realised by several groups  \cite{Broome:2013p7136, Spring:2013p7137, Tillmann:2013p10461, Spagnolo:2014p10480}. 
There have been several theoretical advances in the field of Boson sampling that has extended the protocol to other quantum states of light  \cite{ Lund:2014p10967, Barkhofen:2017p13761, Hamilton:2017p15310, Chabaud:2017p15337, Quesada:2018p17915, Kruse:2019p23056, DeshpandeSciAd22, spagnolo2023non}. 
Recent work examined sampling from general bosonic states  \cite{ChabaudPRL130, ChabaudPRR3}, concluding that the complexity of sampling from the state depends upon the stellar rank of the input state and measurement scheme \cite{Chabaud20_stellarrank}.  

Here we show that this is not the case by analysing Boson sampling from general single-mode input states using a different state generation method than previous work. We can map this problem to a related Gaussian-state Boson sampling problem, allowing us to derive a formula that can be used to calculate photon statistics from such an array of any single-mode input states. 
The time-complexity of this formula (number of terms to be summed to calculate it)  scales as $O(2^R)$ where $R$ is a constant independent of the state, and thus not depending upon the particular nature of the state or its non-classical attributes.

\begin{figure}
\centering
\includegraphics[scale=0.2, trim= 8cm 5cm 3cm 3cm]{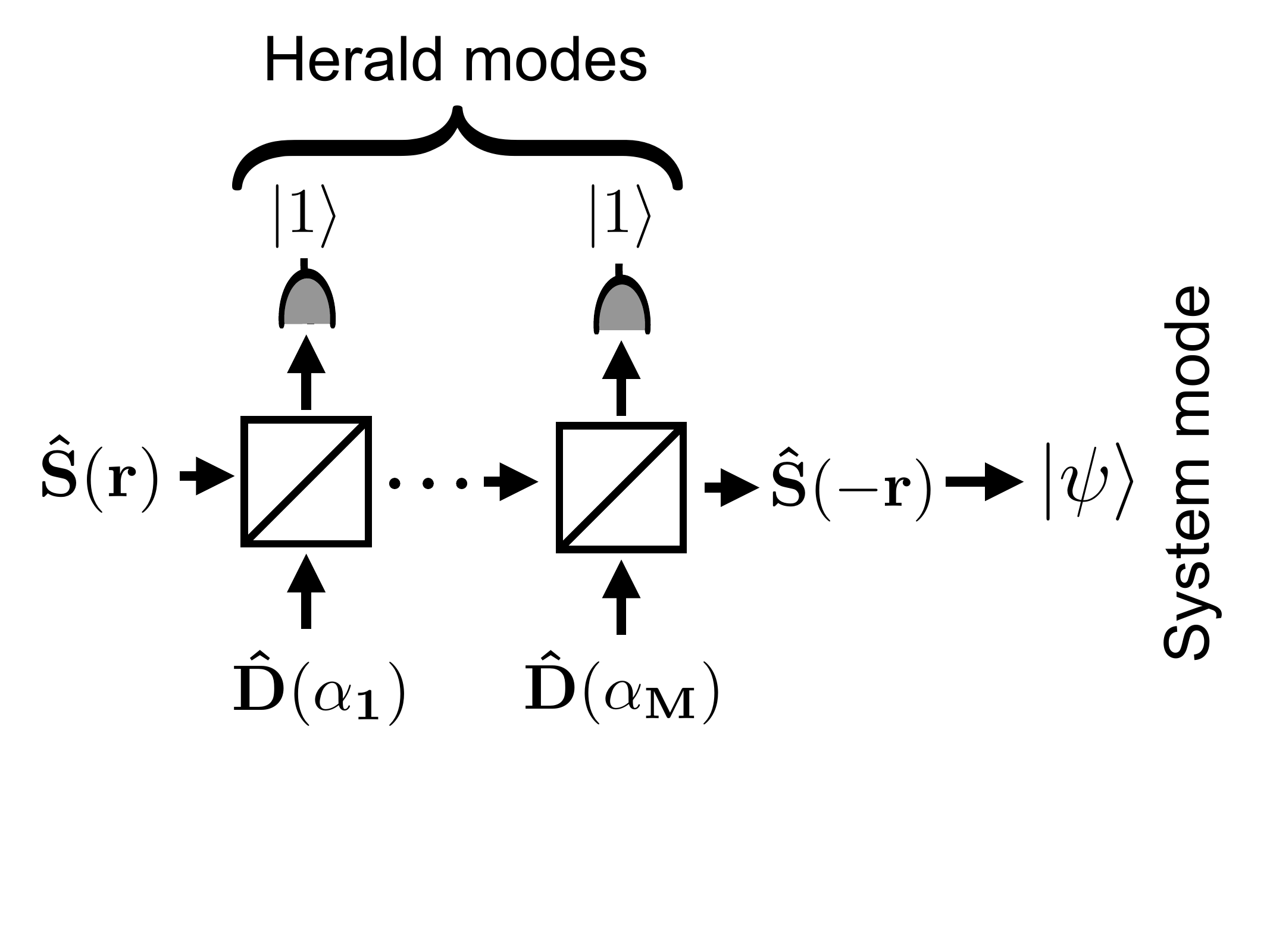}
\caption{Schematic of  non-Gaussian state generation \cite{Fiurasek:2005p20518}. A squeezed vacuum state is generated in the system mode by $\hat{S}(r)$, then travels through a series of high-transmission beamsplitters where it is combined with displaced vacuum states generated by $\hat{D}(\alpha_j)$. After the final beamsplitter the inverse squeezing operation is performed. At the output of each herald mode a single photon must be detected to create the desired state. } \label{fig_ngs_scheme}
\end{figure}

{\it Non-Gaussian state creation from initial Gaussian states}- We briefly describe the scheme devised by 
Fiur\'{a}\v{s}ek \etal~\cite{Fiurasek:2005p20518}, to create any pure single-mode photonic state, by using two squeezing operations, multiple displacement operations and single photon detections. 
Starting from an initial vacuum state, the needed sequence of operations is, 
\be
|\psi\rangle \propto \hat{S}(-r)\aa \hat{D}(\alpha_N)   ...   \aa \hat{D}(\alpha_2) \,\aa \,\hat{D}(\alpha_1) \hat{S}(r) |0\rangle, \label{eq1}
\ee
where $\hat{S}(r)$ is a squeezing operation with parameter $r$, $\hat{D}(\alpha)$ is the displacement operation and $\aa$ is the annihilation operator, a non-unitary operation that removes a photon from the state. 
This state creation process can be seen by noting the effect of the displacement and squeezing operators acting upon $\aa$,
\be
D(\alpha)\aa D^\dag(\alpha) = \aa + \alpha  \hspace{0.5cm} S(r)\aa S^\dag(r) =  \cosh r\,\, \aa + \sinh r \,\,\ad,
\ee
which leads to \eqref{eq1} being written as,
\be
|\psi\rangle \propto (\ad + \alpha_N)... (\ad + \alpha_1) |0\rangle = \sum^{N}_{n=0} c_n \hat{a}^{\dag n} |0\rangle
\ee
which for the correct choice of displacement parameters $\{\alpha_j\}$ generates the desired state. 
An algorithm for these displacement parameters was given in \cite{Fiurasek:2005p20518}.
 In general, a $N$-photon state requires $N$ displacement operators and photon removals, with the two squeezing operations being fixed. 
The authors in \cite{Fiurasek:2005p20518} suggest that such a scheme could be physically  realised by a single-mode squeezed state, travelling through a series of high-transmission beamsplitters, where it is combined with displaced vacuum states and on the reflecting mode of each beamsplitter a single photon is detected, with a final anti-squeezing operation at the output, as sketched in figure~\ref{fig_ngs_scheme}. If the single photons are detected, they herald the creation of the desired state in the remaining mode. 
The downside of this scheme is that the probability to generate such a state decays exponentially with the number of photon measurements required to generate the state. Although this makes it impractical for experimental realisations, it allows us to derive a mathematical expression for calculating the output statistics of interference from such states. 
Utilising the above scheme we can relate, in a unified manner, bosonic Non-Gaussian (NG) states generated from other physical systems, such as Kerr-squeezing or Bose-Hubbard terms, to bosonic sampling and draw some general conclusions.

{\it Gaussian Boson Sampling  -} This NG state generation process uses three operations, squeezing, displacement and photon number measurement, that can all be described within the framework of the Gaussian Boson Sampling (GBS) protocol. 
GBS is a further development of the original BS protocol, where the single photon input states were replaced by a general Gaussian state of light, in particular squeezed vacuum states. 
Gaussian states are completely described by their covariance matrix and displacement vector \cite{schumaker1986quantum, Simon:1994p4225} and the input state remains Gaussian after passing through the linear interferometer.

We now describe the main points of GBS.
$K$ single-mode squeezed states (SMSS) with squeezing parameter $r$, are sent into $K$ modes of an $M'$-mode linear interferometer, characterised by a unitary transformation $\hat{U}$ and what emerges at the output is a multimode squeezed state with covariance matrix $\sigma$. 
The effects of displaced vacuum states can be included by specifying a displacement vector that contains the amplitudes of the coherent states in each mode. 
The measurement statistics from multimode Gaussian states can be calculated using phase-space methods \cite{schleich2011quantum, barnett2002methods}, where each mode of the system is represented by two variables $\alpha_j,\alpha^*_j$, 
the state is described by it's Husmi Q-function and the measurement operators (here number operators) using their Glauber-Sudarshan P-functions.
The probability to measure the photon pattern $\bar{n} = [n_1,n_2,...,n_M]$ from an $M-$mode Gaussian state, described by covariance matrix $\sigma$ and displacement vector $d_v$, is given by \cite{Hamilton:2017p15310, Kruse:2019p23056}
\begin{equation} 
\begin{split}
\Pr(\bar{n}) &= \\ &\frac{\exp[-\frac{1}{2}d^\dag_v \sigma_Q^{-1} d_v ] }{\bar{n}!\sqrt{|\sigma_Q|}}\prod^M_{j =1 } \left (\frac{\partial^2}{\partial \alpha_j \partial \alpha^*_j}\right)^{n_j}  e^{\alpha^t_v A \alpha_v + F\alpha_v} \bigg |_{\alpha_j,\alpha^*_j = 0} \label{main_eq},
\end{split}
\end{equation}
where $\alpha_v = [\alpha_1,\alpha_2,...,\alpha^*_1,\alpha^*_M] $, and the matrix/vector in the exponent are defined by, 
\be
A = \begin{pmatrix} 0_{} &\mathds{I}_M \\ \mathds{I}_M & 0_{}  \end{pmatrix} \left[\mathds{I}_{2M}-\sigma_Q^{-1} \right] \hspace{0.25cm} \mbox{and} \hspace{0.25cm} F = d^\dag_v \sigma_Q^{-1} \label{af_equ},
\ee
with $\sigma_Q=\sigma+\mathds{I}_{2M}/2$, and $\bar{n}! = n_1! n_2!..n_M!$.
If $d_v=0$ then this quantity can be related to the Hafnian \cite{caianiello1973combinatorics} of the matrix $A_S$, which is a submatrix of $A$ that depends upon where the photons are detected, then 
\be
\Pr(\bar{n}) = \frac{1}{\bar{n}!\sqrt{|\sigma_Q|}}\haf(A_s).
\ee
When $d_v\ne0$ the expression in equ.~(\ref{main_eq}) has been termed the loop Hafnian \cite{bjorklund2019faster, Quesada:2019p20751}, 
\be
\Pr(\bar{n}) = \frac{\exp[-\frac{1}{2}d^\dag_v \sigma_Q^{-1} d_v ] }{\bar{n}!\sqrt{|\sigma_Q|}} \lhaf(A_s,F_s), \label{loophaf_equ}
\ee
where $F_s$ is sub-vector that, like $A_s$, depends upon the location of the measured photons. 
These functions are both in the \# P complexity class and hence can form a Boson sampling protocol. 

\begin{figure}
\centering
\includegraphics[scale=0.18, trim= 0cm 6cm 3cm 0cm]{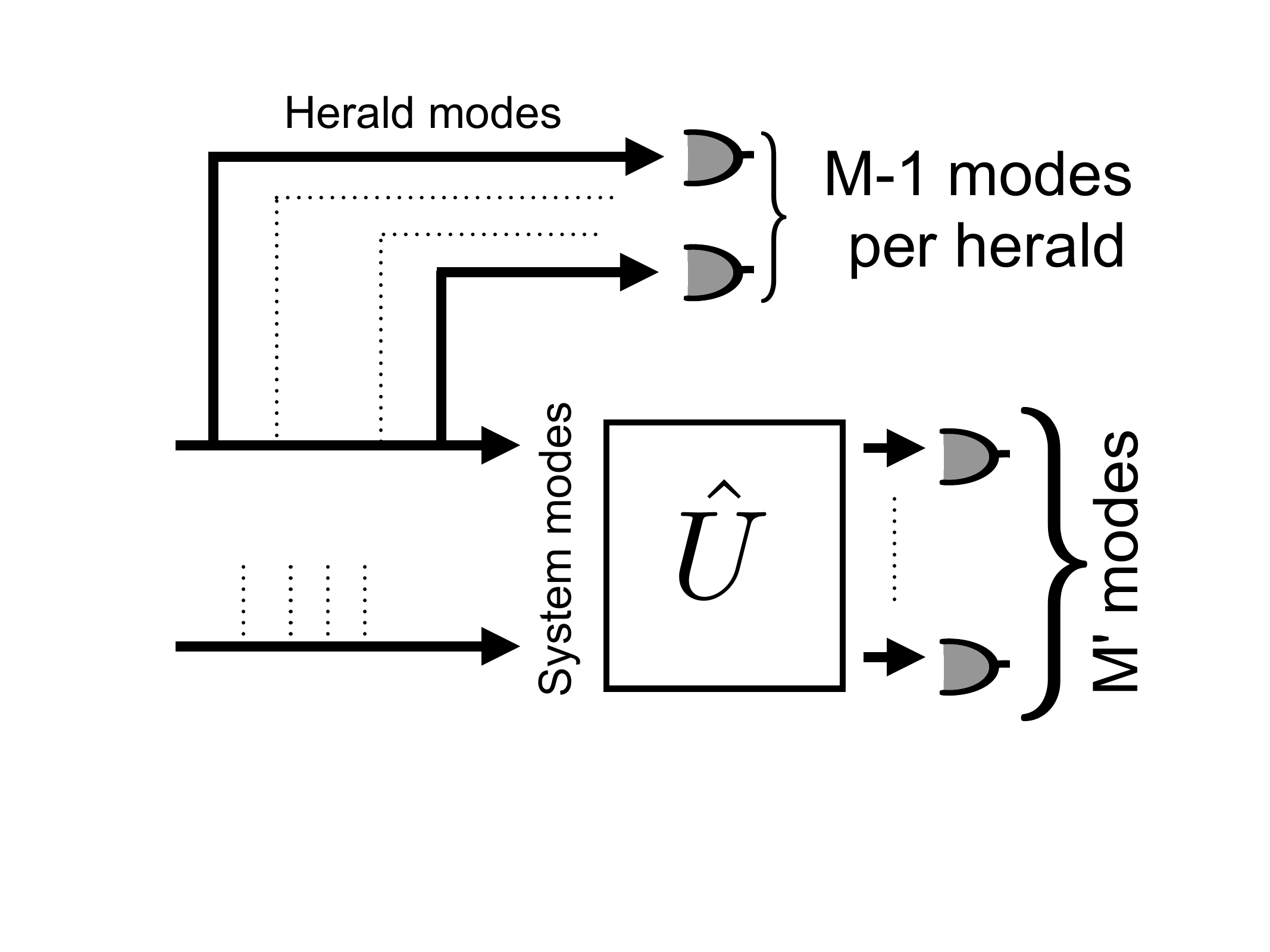}
\caption{Schematic of total sampling from the initial NG states. Each of the input states have $M-1$ herald modes used to generate the remaining state in the system mode, which then enter a linear interferometer. } \label{fig_total_scheme}
\end{figure}

{\it Overall NGBS scheme  -} We now combine the NG state generation scheme with GBS to create a Non-Gaussian Boson sampling (NGBS) problem. 
We will consider the case where we sample from $K$ identical NG states (in principle all input states could be different). 
Each of the input states starts with $M$ modes, of which $M-1$ are herald modes, where the photon detection occurs, and the final system mode that will enter the $M'\times M'-$mode interferometer, which itself have at least $K$ modes. 
Thus the total number of modes is $K(M-1)+M'$. 
The scheme for the NGBS is sketched in figure~\ref{fig_total_scheme}.

The $K$ independent input states can be described by the individual covariance matrices $\sigma_{NG}$ and displacement vectors $d_{NG}$ (before any measurements on the herald modes take place).
The covariance matrix that describes the total input state is the direct sum of the $K$ single-mode covariance matrices and the remaining vacuum modes will be the identity matrix,
\be
\Sigma_{in} = \sigma_{NG}^{\oplus K} \oplus \mathds{I}/2_{M'-K},
\ee
and the total displacement vector is similarly constructed,
\be
\Delta_{in} = d_{NG}^{\oplus K} \oplus 0_{M'-K}.
\ee
where vacuum modes here are zero. 
The system modes of this state then enter the interferometer (described by $U^{sys}_{M'}$) where the output covariance matrix and displacement vector of the total state are easily calculated as,
\be
\begin{split}
&\Sigma_{out} = (U^{sys}_{M'} \oplus \mathds{I}^{herald}_{(M-1)K} ) \Sigma_{in} (U^{\dag\, sys}_{M'} \oplus \mathds{I}^{herald}_{(M-1)K} ) \\  
&\Delta_{out} = (U^{sys}_{M'} \oplus \mathds{I}^{herald}_{(M-1)K} ) \Delta_{in}.
\end{split}
\ee
Finally, every mode, system and herald, is measured in the photon number basis.
This combined scheme can be analysed in terms of GBS in a rather straightforward way as given below.

The NGBS problem is to calculate the conditional probability of the output photon pattern $\bar{n}$ of the interferometer modes, given the $K\times$ NG input states. 
This can be directly related to the GBS problem to measure the photon 
pattern $\bar{n}$ in the interferometric system modes and the pattern $\one$ in the herald modes i.e. single photons. 
Using Bayes' theorem we can relate this conditional probability to the joint probability to measure ${\bar{n}_{system} \cap \one_{herald}}$ in all the modes, divided by the probability to measure $ \one_{herald}$ in the herald modes only,
\be
\Pr(\bar{n}_{system} \big| \one_{herald}) = \frac{\Pr(\bar{n}_{system} \cap \one_{herald})}{[\Pr(\one_{herald})]^K}.
\ee
%\red{Remove}
%\be
%%   \Pr(\bar{n}_{system} \big | \hat{\rho}_{NG}) = \Pr(\bar{n}_{system} \big| \one_{herald})
%%\left (\frac{\partial^2}{\partial \alpha_j \partial \alpha^*_j}\right)^{n_j}  e^{\alpha^t_v B \alpha_v } \bigg |_{\alpha_j,\alpha^*_j = 0}  = 
%\left | \prod^M_{j=1} \left (\frac{\partial}{\partial \alpha_j }\right)^{n_j}  e^{\alpha^t_v B \alpha_v + F\alpha_v} \bigg |_{\alpha_j = 0}  \right |^2
%%\Pr(\bar{n}_{system} \big| \one_{herald}) = \frac{\lhaf(A^{tot}_s, F^{tot}_s)}{p^K}
%%\propto p^{-K} \lhaf( A^{tot}_s, F^{tot}_s  )  %=  \lhaf( p^{-2K/N} A_s, p^{-K/N} F_s  )
%%= |\haf(B)|^2
%\ee
%
%
The probability $\Pr(\bar{n}_{system} \cap \one_{herald})$ can be calculated from (\ref{loophaf_equ}), using covariance matrix $\Sigma_{out}$ and displacement vector $\Delta_{out}$,
\be
\Pr(\bar{n}_{system} \cap \one_{herald}) = \frac{\exp[-\frac{1}{2}\Delta^\dag_{out} \Sigma_Q^{-1} \Delta_{out} ] }{\bar{n}_{sys}!\sqrt{|\Sigma_Q|}} \lhaf( A^{tot}_s, F^{tot}_s  ),
\ee
with $\Sigma_Q  = \Sigma_{out} + \oplus \mathds{I}/2_{M'-K}$ and $A^{tot}_s,F^{tot}_s$ are derived according to (\ref{af_equ}) from $\Sigma_{Q}$ and $\Delta_{out}$ respectively.
Note that all modes of the the total system are measured. The nature of the scheme means that not all rows/columns of $A_s$ are random, due to the enforcement of the herald pattern.

The probability to create the NG input state can be written as,
\be
\Pr(\one_{herald}) =  \frac{\exp[-\frac{1}{2} d^\dag_{NG, s} \sigma_{Q, NG,s}^{-1} d_{NG,s } ] }{\sqrt{|\sigma_{Q, NG,s}|}} \lhaf(A^{NG}_s,F^{NG}_s),
\ee
where $\sigma_{NG,s}$ and $d_{NG,s}$ are the sub-matrix (-vector) formed by deleting the system mode (because it is not measured) and $A^{NG},F^{NG}$ are again derived  according to (\ref{af_equ}).
Using a result by \cite{bjorklund2019faster}, we can absorb the denominator factor, $\Pr(\one_{herald})=p$,  into the numerator's Loop hafnian,
\be
 p^{-K} \lhaf( A_s, F_s  )  =  \lhaf( p^{-2K/N} A_s, p^{-K/N} F_s  ),
\ee
where $N= |\bar{n}+K(M-1)|$, the number of detected photons, including the herald photons, which is also the dimension of $A_s$.
This equation to calculate the statistics from an array of NG states is the main result of this paper and shows that sampling from a NG input state is identical to sampling from a squeezed, displaced vacuum state, where some of the measured output results are fixed (as single photons).

{ \it Complexity of Non-Gaussian Boson Sampling- } It has been shown that the rank of a matrix, $R$, is important for the complexity of the calculation of both  it's  Hafnian and Permanent  \cite{barvinok1996two, bjorklund2019faster}. 
This is due to the number of terms to be summed in each matrix function scales as $2^R$, rather than $2^N$ (where $N$ is the dimension of the matrix) which can be substantially faster for large, low rank matrices. 
In the original BS, the rank of the matrix sampled was equal to number of unique modes that single photons enter and exit, and in GBS the rank of the matrix sampled depends upon the number of unique input squeezed modes  and modes where photons are detected \cite{Hamilton:2017p15310, Kruse:2019p23056}.

Here, we provide evidence that the rank is also important to the time-complexity of the loop Hafnian calculation.
If we examine eq.~\ref{main_eq}, where $A$ is a low rank matrix, we can transform from the $\{\alpha_j,\alpha^*_j\}$ basis to one where $A$ is diagonal and thus we only have $R$ variables, $\{\beta_j,\beta^*_j\}$.
This means that the exponential function is now a product of $R$-single variable functions, whose derivatives are easy to calculate.  
The complexity of the calculation is now that the product of partial derivatives needs to be transformed to this new basis. This is equivalent to expanding a multivariate polynomial \cite{barvinok2016combinatorics},
\be
\prod^N_{j=1} \frac{\partial}{\partial \alpha_j} = \prod^N_{j=1} \left ( \sum^R_{k=1} T_{j,k} \frac{\partial}{\partial \beta_j}   \right )
\ee
whose coefficients in the new basis depend upon the permanent of a matrix derived from the transformation matrix $T$.
Expanding the polynomial in this new basis can be done in $O(N^{R-1})$ operations \cite{barvinok1996two}, i.e. it is dependent upon the rank of the original matrix $A$. Also, each term in the polynomial, i.e $\frac{\partial^{n_j}}{\partial^{n_j} \beta_j} \frac{\partial^{n_j}}{\partial^{n_j} \beta_j} ...\frac{\partial^{n_j}}{\partial^{n_j} \beta_j} $ has a coefficient that depends upon a permanent constructed from the matrix $T$,  that has at most rank $R$, and thus can be calculated in $O(N^{R-1})$ steps \cite{barvinok2016combinatorics}.

This means that the complexity of sampling from these NG states is saturated and does not grow, even as other measures of quantum-ness do, such as negativity of the Wigner function \cite{Kenfack_2004}.
A reason for this saturation is that even though the input states grow in quantum-ness, the measurement basis remains fixed in the number states, and due to the symmetry of quantum mechanics we can consider these as in the input states and measurement in the NG basis. 
Thus there is no immediate advantage to using complicated, general multi-photon single-mode input states to a Boson sampling protocol over simpler input states, such as single photon states or squeezed Gaussian states.

{\it Fidelity of output states - } The authors in \cite{Fiurasek:2005p20518} show that the NG state generation scheme has a fidelity that approaches unity as the transmission coefficient of the beamsplitters approach 1.   
However, we may have to approximate an infinite dimensional state by a finite truncation in the photon number basis. Then, the created state will have a fidelity below unity regardless of any parameters chosen, which may be a source of error in the protocol presented here. 
This can be countered in Boson sampling, as we are generally only concerned with measuring a total of $N$ photons from our output state, thereby truncating our state to this subspace. 
Thus each individual input state only has to be expanded up to this $N$ photon limit and as long as we ensure unit fidelity of this subspace in our input state, then our overall fidelity of the Boson sampling protocol can be unity.

{\it Conclusions- } In conclusion, we have introduced a way to analyse Boson sampling from non-Gaussian states by relating them to Gaussian input states and conditional measurements. 
Thus we arrive at a formula that can be used to calculate the output probabilities of boson numbers from non-Gaussian states entering linear interferometers.
One consequence of this approach is that all pure single-mode states have the same time-complexity scaling that is independent of the particular input state.

This work can extend the photonic sampling problem to other physical bosonic systems that can create more exotic input states than single photons or squeezed light, from, say, higher-order Hamiltonians. 
In future it would be useful to look at two-mode non-Gaussian states, which have a richer structure that single-mode states and may lead to more complex states over single-mode states.

\textbf{Acknowledgements: }

C.~S.~H. and I.~J. received support from the Ministry of Education RVO 68407700 and ÒCentre for Advanced Applied SciencesÓ, Registry No.~CZ.02.1.01/0.0/0.0/16\_019/0000778, supported by the Operational Programme Research, Development and Education, co-financed by the European Structural and Investment Funds and the state budget of the Czech Republic and GA\v{C}R: 23-07169S. 

\bibliography{Hamilton_NGBS.bib}
\bibliographystyle{unsrt}

\end{document}